\documentclass[preprint2]{aastex}
\usepackage{color}

\shorttitle{Solar Fast drifting Radio Bursts} \shortauthors{Tan et al.}

\begin{document}

\title{Solar Fast Drifting Radio Bursts in an X1.3 Flare on 2014 April 25}

\author{Baolin Tan\altaffilmark{1, 3}, Nai-hwa Chen\altaffilmark{2}, Ya-hui Yang\altaffilmark{2}, Chengming Tan\altaffilmark{1, 3}, Satoshi Masuda\altaffilmark{4}, Xingyao Chen\altaffilmark{1}, H. Misawa\altaffilmark{5}}

\affil{$^{1}$CAS Key Laboratory of Solar Activity, National Astronomical Observatories of Chinese Academy of Sciences, Datun Road 20A, Chaoyang District, Beijing 100012, China; bltan@nao.cas.cn}

\affil{$^{2}$Graduate Institute of Space Science and Engineering, National Central University, Taoyuan, Taiwan}

\affil{$^{3}$School of Astronomy and Space Sciences, University of Chinese Academy of Sciences, Beijing 100049, China.}

\affil{$^{4}$Institute for Space-Earth Environmental Research, Nagoya University, Nagoya 464-8601, Japan}

\affil{$^{5}$Planetary Plasma $\&$ Atmospheric Research Center, Tohoku University, 6-3 Aoba, Aramaki, Aoba-ku, Sendai 980-8578, Japan}

\begin{abstract}

One of the most important products of solar flares are nonthermal energetic particles which may carry up to 50\% energy releasing in the flaring processes. In radio observations, nonthermal particles generally manifest as spectral fine structures with fast frequency drifting rates, named as solar fast drifting radio bursts (FDRBs). This work demonstrated three types of FDRBs, including type III pair bursts, narrow band stochastic spike bursts following the type III bursts and spike-like bursts superimposed on type II burst in an X1.3 flare on 2014 April 25. We find that although all of them have fast frequency drifting rates, but they are intrinsically different from each other in frequency bandwidth, drifting rate and the statistical distributions. We suggest that they are possibly generated from different accelerating mechanisms. The type III pair bursts may be triggered by high-energy electron beams accelerated by the flaring magnetic reconnection, spike bursts are produced by the energetic electrons accelerated by a termination shock wave triggered by the fast reconnecting plasma outflows impacting on the flaring looptop, and spike-like bursts are possibly generated by the nonthermal electrons accelerated by moving magnetic reconnection triggered by the interaction between CME and the background magnetized plasma. These results may help us to understand the generation mechanism of nonthermal particles and energy release in solar flares.

\end{abstract}

Online-only material: color figures

\section{Introduction}

In physics of solar flares and coronal mass ejections (CMEs), radio observations play a key role to understand the primary energy release, triggering mechanism of eruptions, and the related plasma instabilities. It can provide the most sensitive direct evidences of magnetic reconnections, nonthermal particle accelerations and propagations, and the variations of magnetic field in corona (Dulk 1985, Bastian et al. 1998). Especially in broadband dynamic radio spectral observations, there are various kinds of radio bursts, such as radio type I, II, III, and IV bursts, and overlying complex spectral fine structures with timescale of sub-second, including spike bursts (Benz et al. 1982, Tan 2013, etc.), fiber bursts (Chernov et al. 2010), Zebra patterns (Tan et al. 2014), quasi-periodic pulsations (Tan et al. 2010), etc. Different radio bursts and the related fine structures would reflect different physical processes in source region, including different physical conditions, coupling interactions, accelerations, and the different kinetic energy of the nonthermal particles, etc.

Among various kinds of radio bursts, many of them show an important feature: frequency drifting rate $D=\frac{df}{dt}$, which is manifested as a slope of the radio burst pattern on the spectrogram and reflects the motion of the emitting medium. In order to make a reasonable comparison among different kinds of bursts at different frequency ranges, we generally define a relative frequency drifting rate:

\begin{equation}
\bar{D}=\frac{df}{f_{0}dt}.
\end{equation}

Here, $f_{0}$ is the central frequency of the burst. Type II radio burst at meter wavelengths present slow frequency drift rate with $\bar{D}\leq0.01$ s$^{-1}$ which reflects the motion of a CME (a big cloud of fast plasma flow), type III radio burst shows very fast frequency drift rate with $\bar{D}\geq0.1$ s$^{-1}$ and implies the fast flight of nonthermal electron beams in corona, the moving type IV radio burst also shows a very slow frequency drift rate $\bar{D}\ll0.01$ s$^{-1}$ and reveals the motion of the emitting corona loops, etc. Additionally, the frequency drifting rate may have positive or negative sign. The positive drifting (PD) rate means the emission drifting from low frequency to high frequency and the emitting source may move from high place downward to lower place in solar corona conditions. The negative drifting (ND) rate means the emission drifting from high frequency to low frequency and the emitting source moves from low place upward to higher place (fly out from the solar surface up to high corona). Therefore, solar radio bursts can be classified into two types: PD bursts and ND bursts.

Generally, according to the magnitude of $\bar{D}$, solar radio bursts can be classified into another two categories,

(1) Slow drifting radio burst (SDRB). Its frequency drifting rate is very slow ($\bar{D}<0.01$ s$^{-1}$ in most cases) and the corresponding moving velocity of the emitting medium is near or slower than the local Alfven speed ($v_{A}$). They are possibly produced by some plasma flows, jets or the motions of coronal loops. Such as type II radio bursts (Dulk 1985) and moving type IV radio bursts (Dulk \& Altschuler 1971), etc.

(2) Fast drifting radio burst (FDRB). The frequency drifting rate is very fast ($\bar{D}>0.1$ s$^{-1}$ in most cases) and its corresponding moving velocity of the emitting medium is much faster than the local $v_{A}$. Sometimes, the velocity is close to relativistic level. They provide key information of the motion of nonthermal electron beams. Such as type III bursts (Reid \& Ratcliffe 2014), spike bursts (Tan 2013), etc. Actually, type III bursts are believed to be a sensitive signature of the energetic electron beams generated and propagated in the corona (Lin \& Hudson 1971, Lin et al. 1981, Aschwanden et al. 1993, and a recent review in Reid \& Ratcliffe 2014).

FDRBs are related to the motion of energetic electrons, it is very important to study their characteristics for understanding the primary energy release, particle acceleration and the energy transportation in flaring processes (Miller et al. 1997). This work reports three distinctly different groups of FDRBs in a powerful X1.3 flare on 2014 April 25. It is very interesting to show a group of spike bursts just following a group of type III pair bursts around the separate frequency with stochastic distribution, and a group of spike-like bursts superimposed on a type II radio burst. Section 2 presents the observing features of the three kinds of FDRBs, including their possible relationship with hard X-ray emission (HXR), extreme ultraviolet (EUV) bursts and the flare and CME processes. Section 3 presents the discussions of possible physical mechanisms. Main conclusions are summarized in Section 4.

\section{Observations of the solar fast drifting radio bursts}

The related flare occurred at the solar west limb on 2014 April 25 which was fully observed by several telescopes, including soft X-ray (SXR) at 0.5 - 4 \AA~ and 1 - 8 \AA~ observed by GOES, extreme ultraviolet (EUV) images observed by SDO/AIA (Lemen et al. 2012), hard X-ray (HXR) by RHESSI (Lin et al. 2002), radio dynamic spectrogram at frequency of 100 - 500 MHz by Iitate Planetary radio telescope (IPRT/AMATERAS, Iwai et al. 2012) and microwave images at frequency of 17 GHz observed by Nobeyama Radio Heliograph (NoRH), etc.

The SXR light curve shows that the flare starts at 00:17 UT, peaks at 00:25 UT, and ends at 00:45 UT with a magnitude of X1.3 class in NOAA Active Region 12035 located just behind the solar limb. In the flare impulsive phase, a lower arcade takes off as a strong expansion CME at about 00:22 UT with averaged speed of 600 km s$^{-1}$ (Chen et al. 2016). Accompanying to the flare, several groups of radio bursts are observed, which will be presented in the following sections.

\subsection{Main characteristics of the fast drifting radio bursts}

Fig. 1 presents the solar radio dynamic spectrogram at frequency of 100 - 500 MHz observed by IPRT/AMATERAS overplotted SXR flux at 1 -8 \AA (white solid line) observed by GOES and HXR light curves (green:12-25 keV, blue: 25-50 keV, yellow: 50-100 keV) observed by RHESSI.

The HXR 50-100 keV and 25-50 keV light curves reach to maximum around 00:20 UT, about 5 minutes before the SXR peak time, which are co-temporal with the onset of a group of type III radio bursts. The flare light curve shows a sharp and smooth rising phase and a long smooth decay phase in which the impulsive phase can be defined as the FWHM of HXR 50-100 keV light curve, e.g. about 00:20:00 UT to 00:23 UT.

\begin{figure*} % Fig. 1
\begin{center}
   \includegraphics[width=16.0 cm]{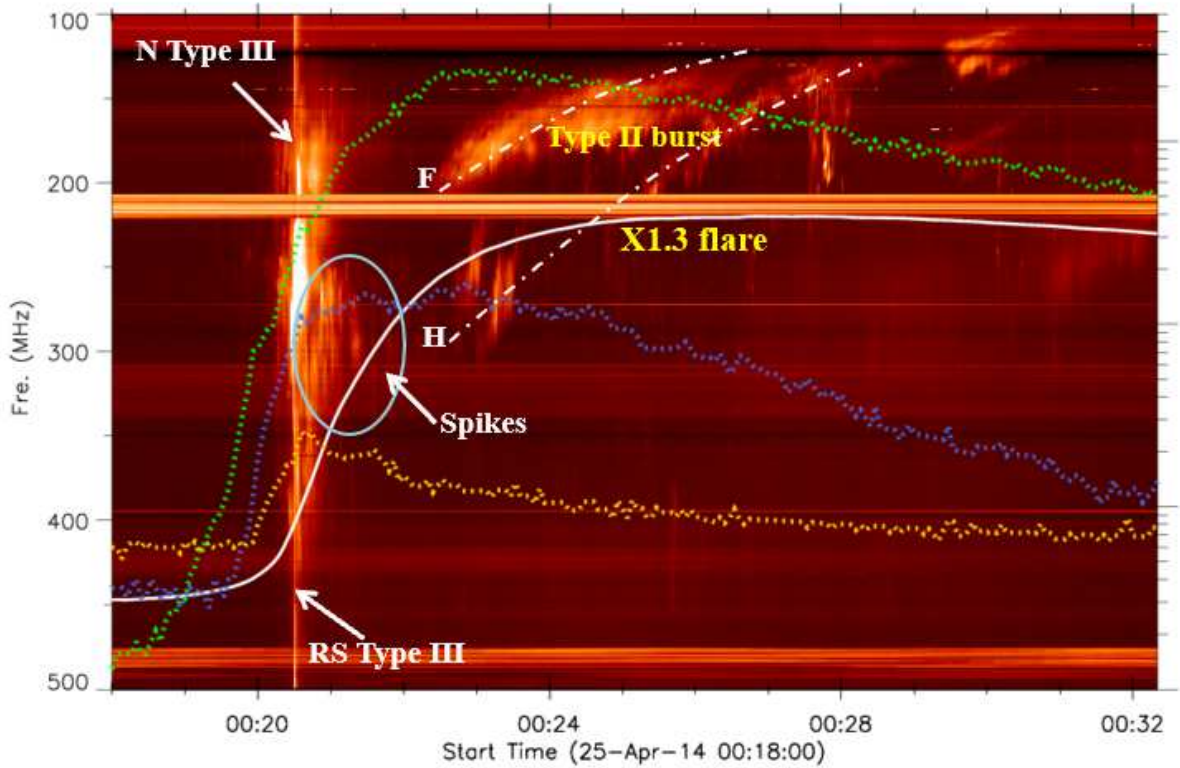}
\caption{Radio spectrogram observed by IPRT/AMATERAS in an X1.3 flare on 2014 April 25. It shows a group normal type III bursts (N type III) and a group reverse-sloped type III bursts (RS type III) form a group type III pair bursts at the flare start time. A group spike bursts (in the blue ellipse) are following the type III pair bursts around the separate frequency. Then, a type II burst occurs with fundamental (F) and harmonic (H) branches and superimposed by a group spike-like bursts. The white curve is the SXR flux observed by GOES. The dashed green, blue, and yellow curves show the emission counts of HXR at energy 12-25 keV, 25-50 keV, and 50-100 keV, respectively. }
\end{center}
\end{figure*}

IPRT/AMATERAS is a solar radio spectral metric polarimeter at the Iitate observatory in Fukushima prefecture, Japan. Its minimum detectable flux is less than 0.7 sfu with cadence of 10 ms and frequency resolution of 61 kHz. Both left and right circular polarizations are simultaneously observed (Iwai et al. 2012). With such high performance, we have identified several groups of FDRBs around the flare: a group of type III pair bursts (N type III and RS type III), a group of spike bursts just following the type III pair bursts around the separate frequency, and a great group of spike-like bursts superimposing on a type II bursts with fundamental (F) and harmonic (H) branches.

\subsubsection{Type III Pair Bursts}

The type III bursts occur around 00:20:28 UT at frequency range of 100 - 500 MHz. Actually, the type III bursts can be plotted into two groups. One group has negative frequency drifting rates and named as normal type III bursts (N type IIIs) while the other group has positive frequency drifting rates and named as reverse-sloped type III bursts (RS type IIIs).

The N type IIIs have frequency drifting rates from -150 MHz s$^{-1}$ to  -290 MHz s$^{-1}$, composed with 8 type III bursts. $\bar{D}$ is in the range of from -0.85 s$^{-1}$ to -1.25 s$^{-1}$. They start at frequency of about 310 MHz and extend down to below 100 MHz (lower limit of the instrument). The duration of each single normal type III burst is in the range of 0.2 - 0.6 s with average of 0.41 s. The frequency bandwidth $f_{w}$ is from 45 MHz to about 110 MHz with average of 75 MHz, the relative bandwidth $\bar{f}_{w}=\frac{f_{w}}{f_{0}}$ ranges from 0.25 to 0.53, and the average is about 0.38.

The RS type III bursts have positive frequency drifting rates from 230 MHz s$^{-1}$ to 370 MHz s$^{-1}$, composed with 7 type III bursts, $\bar{D}\approx$ 0.70 s$^{-1}$ to 1.29 s$^{-1}$. They start at frequency of about 330 MHz and extend up to beyond 500 MHz (the upper limit of the instrument). The duration of each single RS type III bursts is about 0.3 - 0.7 s with average of 0.48 s. The frequency bandwidth $f_{w}$ is from 50 MHz to 120 MHz with average of 89 MHz, and $\bar{f}_{w}$ ranges from 0.29 to 0.55 with average of 0.41.

It is well accepted that type III radio bursts belong to the category of FDRB which are associated to some nonthermal electron beams. The separate frequency ($f_{x}$) between the normal and RS type III bursts is about 310 MHz and the frequency gap is around 20 - 25 MHz. The two group bursts form a group of radio type III pairs (Robinson \& Benz 2000, Ning et al. 2000). The whole group of type III pairs lasts for about 5 s.

Fig. 1 shows that the type III pairs just appear around the peak time of high energy HXR light curve at 50 - 100 keV and fast rising phase of HXR emission at 12 - 25 keV and 25 - 50 keV. \textbf{Imaging observation (Fig. 6)} shows that HXR source and microwave source at 17 GHz are locate around a same cusp-shaped structure. Chen et al. (2016) proposed that a breakout magnetic reconnection and particle acceleration are occurred just around this magnetic configuration. The co-spatial and co-temporal relationship between HXR source, microwave source and cusp-shaped structures indicate that the type III pair bursts should be possibly associated to the energetic electron beams with energy of 50 - 100 keV. These energetic electrons should be accelerated in the above magnetic reconnections.

\subsubsection{Spike Bursts Following the Type III Pair Bursts}

Just following the type III pair bursts, there are a group radio spike bursts occurring from 00:20:30 UT to 00:21:25 UT at frequency range of 220 - 410 MHz. Fig. 1 shows that the radio spike bursts are deviate from the extension zone of the type II burst. They are locating very close to the separate frequency ($f_{x}$) of the type III pair bursts and around the peak time of the HXR at 25 - 50 keV, the rapid rising time of HXR at 12 - 25 keV and the slowly decay time of HXR at 50 - 100 keV. This fact implies that the spike bursts are most possibly relevant to the magnetic reconnection, but the related nonthermal electrons should have relatively lower energy than that of the above type III pair bursts.

\begin{figure*} % Fig. 2
\begin{center}
   \includegraphics[width=15.0 cm]{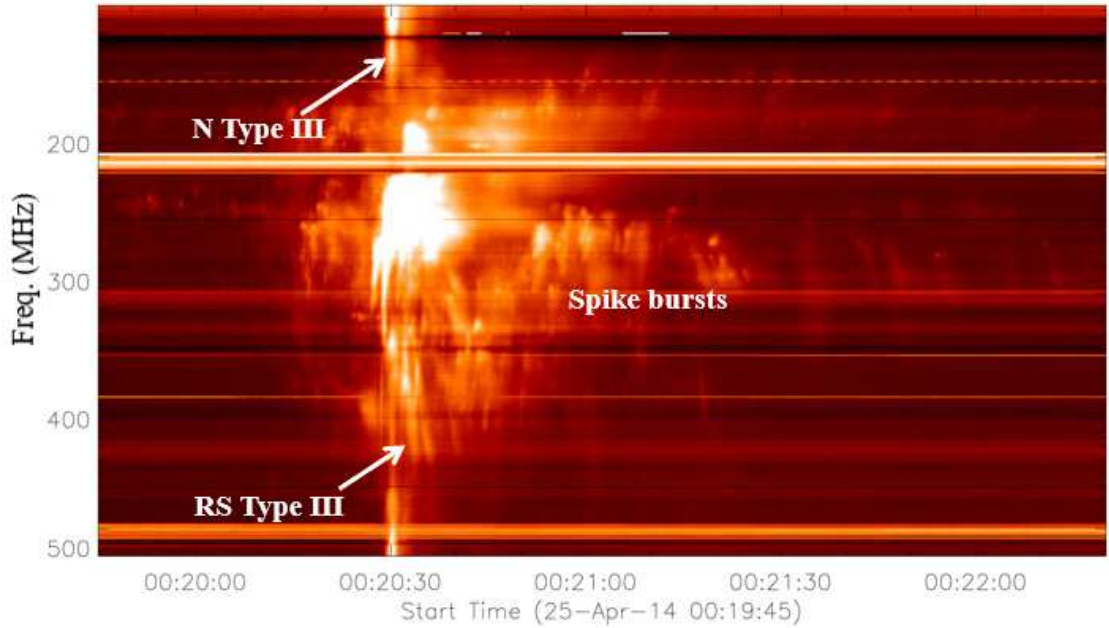}
\caption{Enlargement of the radio dynamic spectrogram of type III pair bursts and the following spike bursts.}
\end{center}
\end{figure*}

Fig. 2 presents the enlarged spectrum of the type III pair bursts and the following spike bursts during the flare impulsive phase. The spike bursts are distributed randomly on the time-frequency spectrogram. We perform a figure auto-recognition and quantify these spike bursts with parameters: central frequency ($f_{0}$), frequency bandwidth ($f_{w}$), lifetime ($\tau$), and frequency drifting rate ($D$). Here, we recognized 245 spikes during 00:20:30 UT - 00:21:27 UT in the frequency range of 240 - 410 MHz.

The statistical analysis shows that the occurrence rate of spike bursts is about 4.6 spikes per second. Among them, there are about 88 spike bursts (34\%) with positive frequency drifting rates from 26.5 MHz s$^{-1}$ to 66.9 MHz s$^{-1}$, $\bar{D}=$ 0.11 $\sim$ 0.26 $s^{-1}$, named as positive drifting spikes (PD spikes). The averaged positive frequency drifting rate is about 47.5 MHz s$^{-1}$ and the average $\bar{D}$ is about 0.19 $s^{-1}$. The other 157 spike bursts (66\%) have negative frequency drifting rates from -39.2 MHz s$^{-1}$ to -138.2 MHz s$^{-1}$, $\bar{D}=$ -0.13 $\sim$ -0.39 $s^{-1}$, named as negative drifting spikes (ND spikes). The averaged negative frequency drifting rate is about -68.1 MHz s$^{-1}$ and the average $\bar{D}$ is about -0.29 $s^{-1}$. Obviously, the frequency drifting rates of spike bursts are smaller about one order than that of the type III pair bursts. But they are still much faster than that of the type II bursts. The analysis in Section 3 shows that the spike bursts still belong to FDRBs.

\begin{figure*} % Fig. 3
\begin{center}
   \includegraphics[width=12.0 cm]{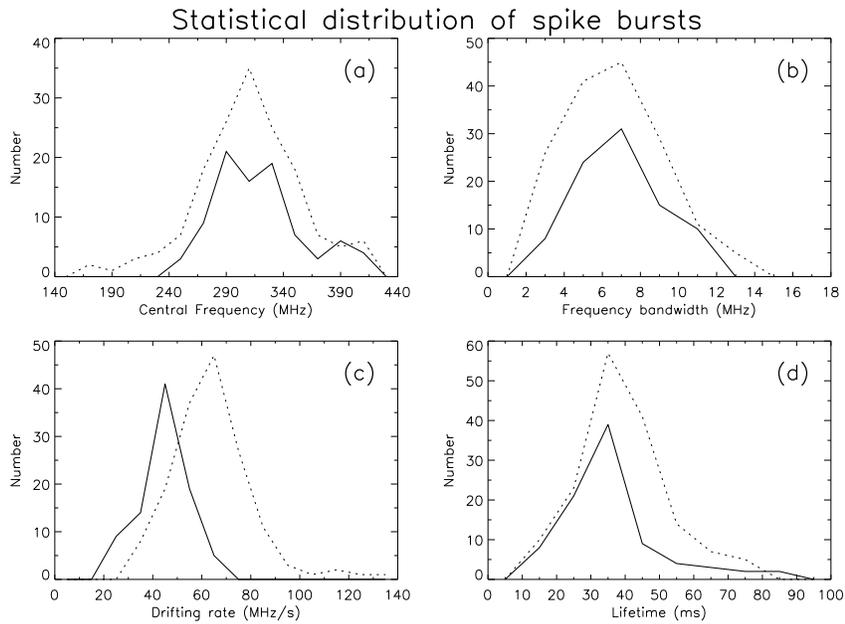}
\caption{The statistical distributions of the central frequency (a), frequency bandwidth (b), frequency drifting rate (c) and lifetime (d) of spike bursts. The solid and dashed lines are expressed the PD and ND spike bursts, respectively.}
\end{center}
\end{figure*}

Fig. 3 presents the statistical distributions of central frequency (a), frequency bandwidth (b), frequency drifting rate (c) and lifetime (d) of spike bursts. It shows that the central frequency of PD and ND spikes range from 250 MHz to 410 MHz and from 170 MHz to 410 MHz, respectively, and their profiles reach to maximum at the almost same frequency. In other word, different from the distributions of normal type III and RS type III bursts, PD and ND spikes are distributed stochastically in the same frequency range. The frequency bandwidths of the PD and ND spikes range from 2.6 MHz to 12.0 MHz and from 3.3 MHz to 14.0 MHz, respectively. The relative bandwidth $\bar{f}_{w}$ ranges from 0.01 to 0.04 with averaged values 6.4 MHz and 6.8 MHz, respectively, only at the order of one or several percents of the central frequency. The frequency bandwidths of spike bursts are narrower at least one order than that of the above type III pair bursts. The lifetimes of each single PD and ND spikes range from 10 ms to 90 ms and from 10 ms to 80 ms, with averaged lifetimes of 35.4 ms and 39.8 ms, respectively. The lifetime is shorter at about one order than that of the type III bursts.

In brief, the spike bursts are distributed randomly following the type III pair bursts around the separate frequency. They are short lifetimes, narrow frequency bandwidth, and mildly frequency drifting rates. Comparing to the type III pair bursts relating to high energy nonthermal electrons, the narrow band stochastic radio spike bursts are possibly related to some relatively low energy nonthermal electrons.

\subsubsection{Spike-like bursts on Type II Burst}

The type II burst occurred from 00:22:20 UT to 00:31:20 UT below the frequency of 280 MHz, about 1 minute after the above spike bursts and about 2 minutes after the type III pair bursts. It is composed of two branches and forms a harmonic structure (Fig. 4). The fundamental branch (F) extends from about 184 MHz down to below 100 MHz, while the second harmonic branch (H) extends from about 286 MHz down to near 100 MHz. The frequency ratio between the fundamental and harmonic branches is about 1.55. The frequency bandwidth of the type II burst strips is in the range of 40 - 80 MHz. The frequency drifting rates of the fundamental and the second harmonic branches are very slow, about 0.3 MHz s$^{-1}$ and 0.5 MHz s$^{-1}$, with $\bar{D}$ of about -0.002 s$^{-1}$ and -0.003 s$^{-1}$, respectively. These rates are about 1 or 2 orders lower than that of the spike bursts, and 3 orders lower than that of the type III pair bursts. Therefore, the type II burst belongs to a category of SDRB.

\begin{figure*} % Fig. 4
\begin{center}
   \includegraphics[width=15.0 cm]{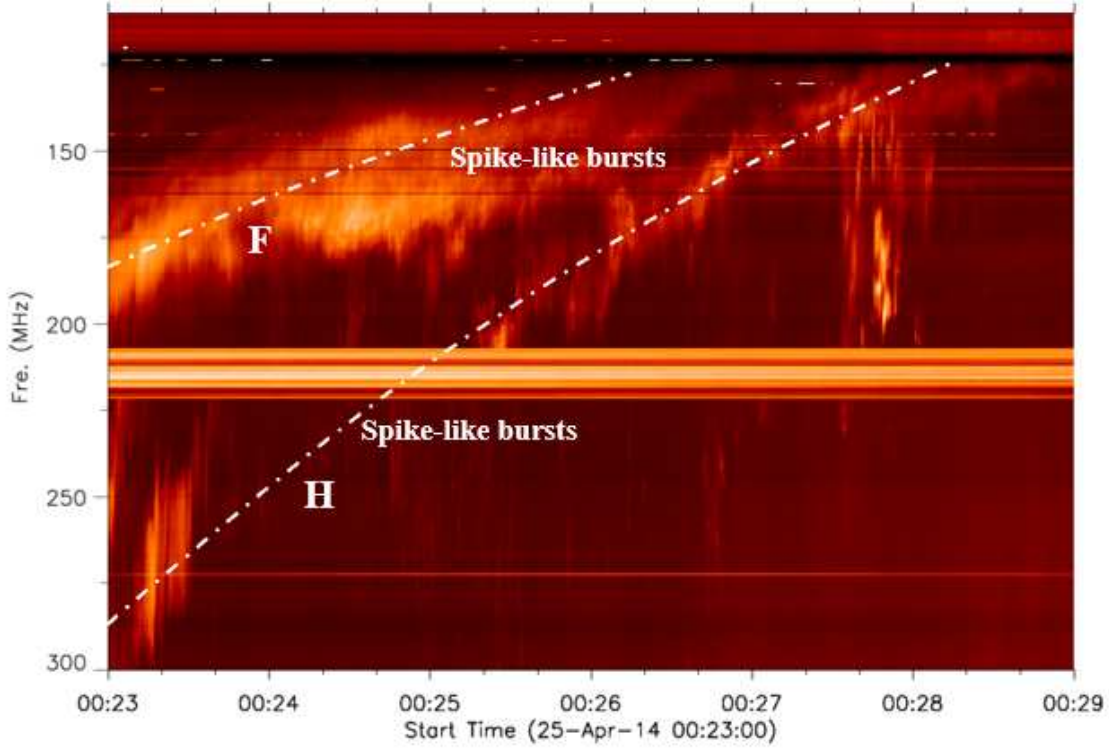}
\caption{The enlargement of the radio dynamic spectrogram of the type II burst and its fine structures of spike-like bursts. The dot-dashed lines marked the ridges of fundamental (F) and harmonic (H) branches of the type II burst.}
\end{center}
\end{figure*}

Similar to most type II radio bursts reported in other literatures (e.g. Armatas et al. 2019, etc.), we found that the type II burst also contain super-fine structures with groups of spike-like bursts. Fig. 4 presents the enlargement of the radio dynamic spectrogram of the type II burst and its fine structures. Here we identified 140 spike-like bursts on the fundamental and second harmonic branches and measured their central frequency, frequency bandwidth, frequency drifting rates and lifetime, respectively. Fig. 5 presents their statistical distribution. We find that frequency bandwidth of each single spike-like burst is in the range of 7 - 25 MHz with averaged value of 12 MHz, the relative bandwidth $\bar{f}_{w}$ ranges from 0.04 to 0.12, near 2 times of the above spike bursts. The lifetime of each single spike-like burst is about 30 - 60 ms with averaged value of about 42 ms, very similar to the spike bursts. As for the frequency drifting rate of each single spike-like burst, 65 of the 140 identified spike-like bursts are negative from -71 MHz s$^{-1}$ to -133 MHz s$^{-1}$ with $\bar{D}$ in the range from  -0.37 s$^{-1}$ to -0.71 s$^{-1}$, while the other 75 spike-like bursts are positive from 83 MHz s$^{-1}$ to 143 MHz s$^{-1}$ with $\bar{D}$ in the range from 0.41 s$^{-1}$ to 0.67 s$^{-1}$. The drifting rates are about 2 times faster than that of the spike bursts and obviously slower than the type III pair bursts. Different from the stochastic distribution of the spike bursts, Fig. 5(a) shows that the ND spike-like bursts are mainly appeared in the lower frequency side of the ridges (the white dot-dashed lines in Fig. 4) of fundamental (F) and harmonic (H) branches of the type II burst, while the PD spike-like bursts are mainly appearing in the higher frequency side of the ridges. This distribution is very similar to that of the type III pair bursts.

\begin{figure*} % Fig. 5
\begin{center}
   \includegraphics[width=12.0 cm]{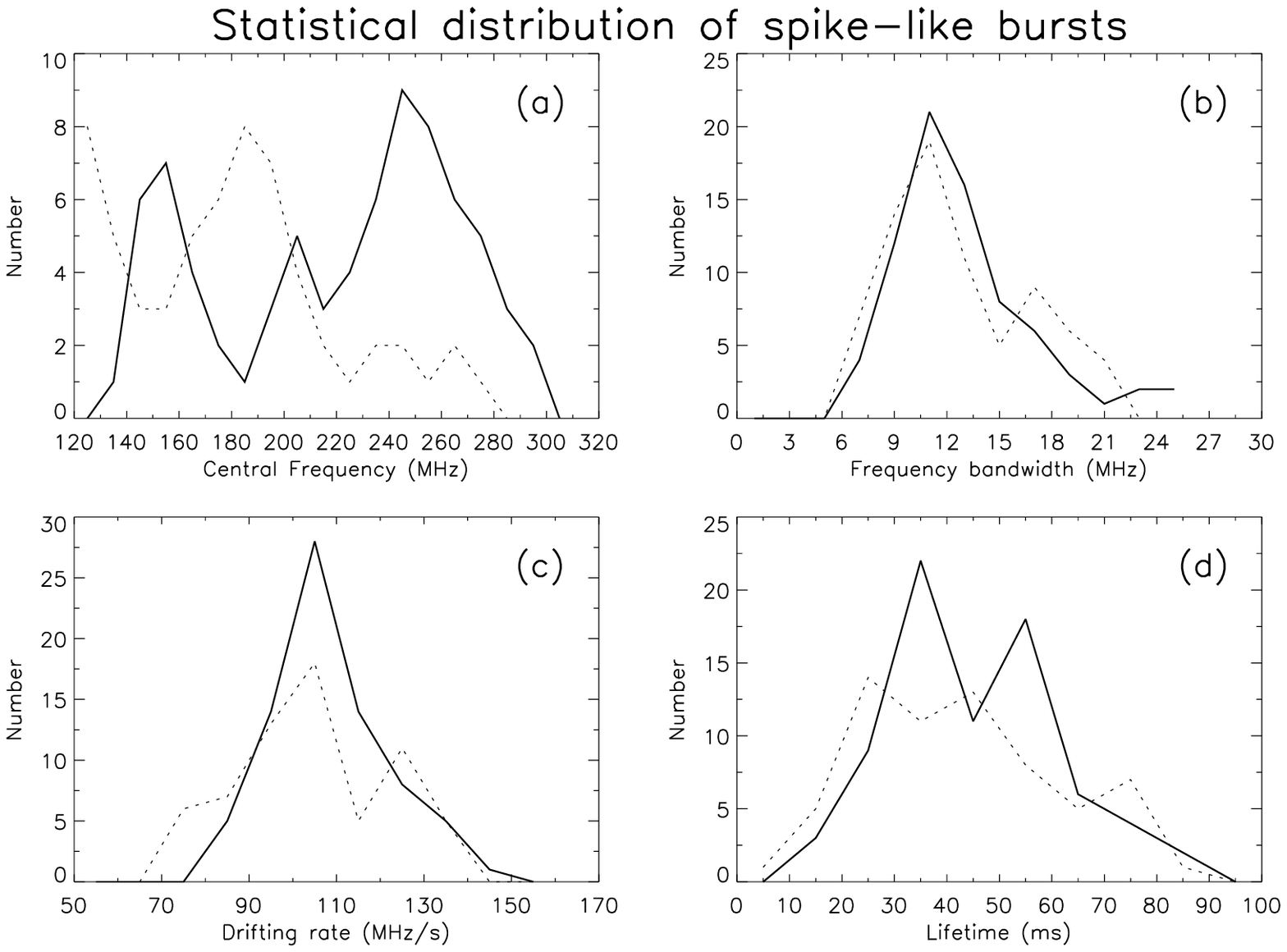}
\caption{The statistical distributions of the central frequency (a), frequency bandwidth (b), frequency drifting rate (c) and lifetime (d) of spike-like bursts on type II burst. The solid and dashed lines are expressed the PD and ND spike-like bursts, respectively.}
\end{center}
\end{figure*}

Chen et al. (2016) reported that during the flare rising phase, a lower arcade above the flaring region takes off as a CME with a rapid expansion and averaged speeds of about 600 km s$^{-1}$. And the above radio type II burst just occurred at the same period, which indicates that both of them may have closely relationship with each other, we will discuss it in Section 3.

\begin{deluxetable}{ccccccccccccccccc}
\tablecolumns{16} \tabletypesize{\scriptsize} \tablewidth{0pc}
\tablecaption{Comparisons among the fast frequency drifting radio bursts  \label{tbl-1}} \tablehead{
\colhead{Burst type} &\colhead{Type III pair burst} &\colhead{Spike burst} &\colhead{Spike-like burst}   \\
}
  \startdata
  Frequency bandwidth (MHz)                  &    45 - 120 (85)    &   2.6 - 14.0 (6.6)   &     7 - 25 (12)     \\
  Relative Frequency bandwidth $\bar{f}_{w}$ &  0.25 - 0.55 (0.42) &  0.01 - 0.04 (0.03)  &  0.04 - 0.12 (0.07) \\
  Lifetime (ms)                              &   200 - 700 (430)   &    10 - 90 (37.6)    &    30 - 60 (42)     \\
  Negative relative drifting rate (s$^{-1}$) &  0.85 - 1.25 (1.03) &  0.13 - 0.39 (0.29)  &  0.37 - 0.71 (0.56) \\
  Positive relative drifting rate (s$^{-1}$) &  0.70 - 1.29 (1.01) &  0.11 - 0.26 (0.19)  &  0.41 - 0.67 (0.53) \\
   \enddata
 \tablecomments{The number in the brackets is the average value of the corresponding parameter. }
\end{deluxetable}

Table 1 lists the comparisons among the three kinds of fast frequency drifting radio bursts with their main parameter characteristics. It shows that spike-like bursts have about 2 times of bandwidth and about 3 times of drifting rates of spike bursts, type III pair bursts have the biggest relative frequency bandwidths and fastest frequency drifting rates, the ND and PD spikes are randomly distributed on the spectrogram which is obviously different from the type III pair bursts and spike-like bursts. Both of normal type III bursts and ND spike-like bursts are mainly distributed in the lower frequency sides, while both of the RS type III bursts and PD spike-like bursts are distributed mainly in the high frequency sides. In brief, there are three kinds of FDRBs: type III pair bursts, spike bursts, and spike-like bursts, which should be intrinsically related to the flare but have different physical processes.

\subsection{Source Region}

As we have no imaging observations at the corresponding radio frequency, it is difficult to determine the location of the source region of radio bursts. However, because the flare is a limb event, we may obtain an estimation of the source region indirectly from multi-wavelength observations.

Fig. 6 shows three consecutive snapshots (every 1 min) at three EUV (171 {\AA}, 131 {\AA}, 94 {\AA}) channels observed by AIA/SDO which present the evolutionary processes during the flare impulsive phase. The overlaid contours are microwave emission at 17 GHz observed by NoRH (white) and HXR emissions observed by RHESSI in the selected time at energy of 6-12 keV (blue), 12-25 keV (yellow), 25-50 keV (red) and 50-100 keV (pink), respectively.

\begin{figure*} % Fig. 6
\begin{center}
   \includegraphics[width=16.0 cm]{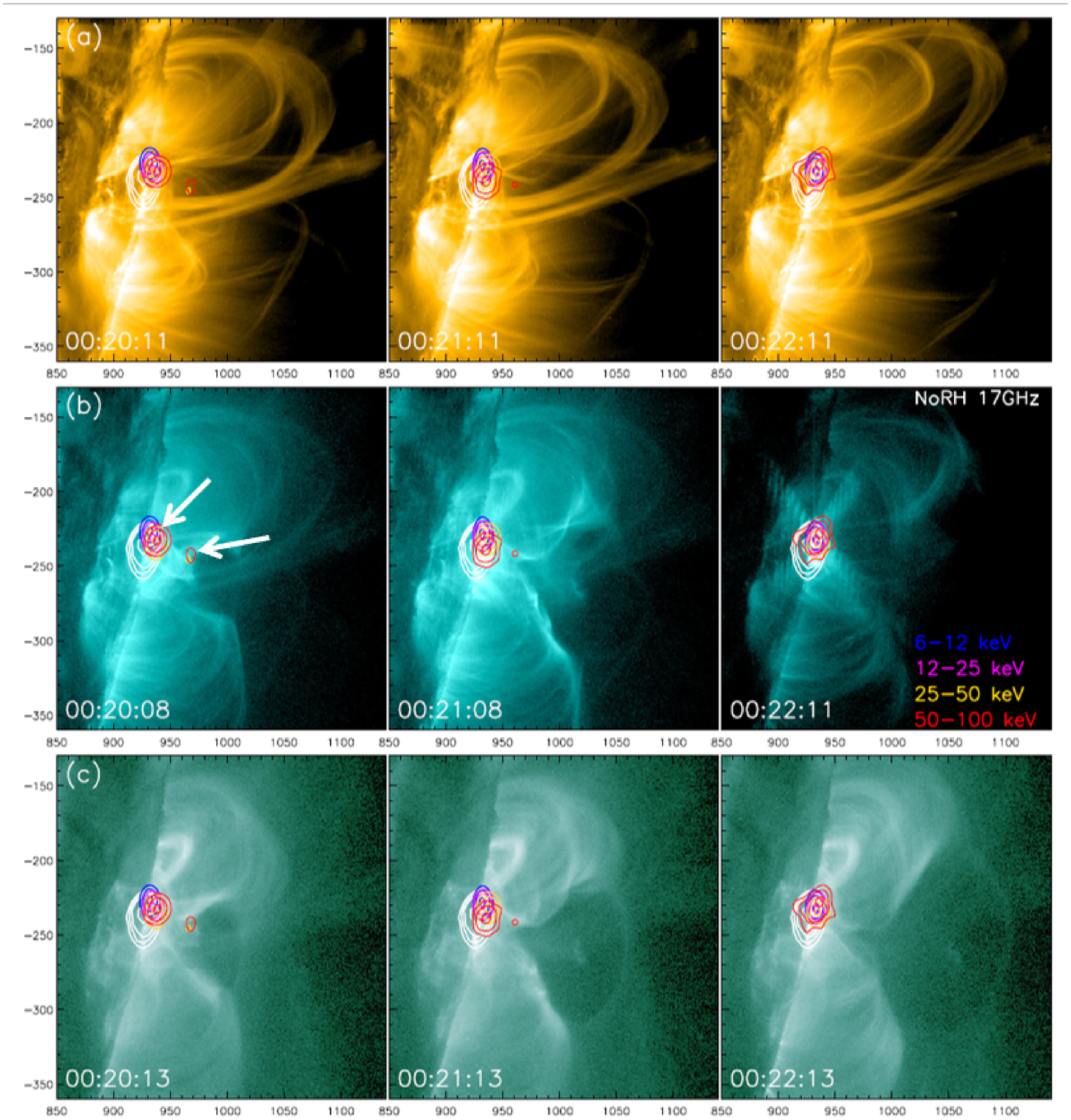}
\caption{The flare consecutive snapshots in every minute at channels of 171 \AA, 131 \AA, and 94 \AA~ in the impulsive phase of the X1.3 flare on 2014 April 25. The overlaid contours are microwave emission at 17 GHz observed by NoRH (white) and hard X-ray emissions observed by RHESSI in the selected time at energy of 6-12 keV (blue), 12-25 keV (yellow), 25-50 keV (red) and 50-100 keV (pink), respectively.}
\end{center}
\end{figure*}

Fig. 6 shows that there is a small compact upper HXR source and a large lower \textbf{HXR} source at energy of 50-100 keV just located very closed to a cusp-shaped structure on the EUV images where brighten sequentially at high temperature emission of 131 \AA~ delineating the newly reconnected field lines around 00:20 UT. The HXR sources are the indicative of strong energy release which could be either the flare footpoints or the site where the magnetic reconnection took place. Two neighboring loop systems (north and south) are involved and the HXR sources are situated between them. At the same time, a microwave burst superimposed almost on the large lower HXR source. A systematical investigation of the flare and the following CME had been demonstrated by Chen et. al. (2016) where suggested that the eruption should be caused by breakout magnetic reconnection around the multi-polarity regions.

We proposed that the small compact upper HXR source should be responsible for the normal type III bursts, while the large lower HXR source for the RS type III bursts. The mid-position between the upper and lower sources should be responsible for the reconnection and acceleration site where the height is about 65 arcsecond ($H=$4.7$\times10^{4}$ km) above the solar surface, and the corresponding radio emission occurred just near the separate frequency (about 310 MHz). Because the spike bursts also occurred very close and around the separate frequency, their source region must be near the above acceleration site. The spike-like bursts on type II burst should be above the upper HXR source with height beyond 70 arcsecond above the solar surface, $H>$5.1$\times10^{4}$ km.

From the mechanism of plasma emission, the plasma density around the reconnection and acceleration site (near the separate frequency) can be estimated about 1.2$\times10^{15}$ m$^{-3}$. The frequency range of spike bursts is 170 - 410 MHz and their corresponding plasma density is from 3.6$\times10^{14}$ m$^{-3}$ to 2.1$\times10^{15}$ m$^{-3}$. And the frequency range of spike-like bursts is 100 - 286 MHz and the plasma density should be from 1.2$\times10^{14}$ m$^{-3}$ to 1.1$\times10^{15}$ m$^{-3}$.

The another parameter is the magnetic field in the source region. The flare event is very close to the solar limb, we have no reliable measurement of the magnetic field. We may indirectly estimate it from a fitted method of Dulk \& MeClean (1978):

\begin{equation}
B=0.5(\frac{r}{R_{s}}-1)^{-\frac{3}{2}}=0.5(\frac{H}{R_{s}})^{-\frac{3}{2}}.
\end{equation}

The unit of magnetic field ($B$) is Gs and $R_{s}$ is the solar photospheric radius. $r$ is the distance from source region to solar center. $H$ is the height above solar photospheric surface. Then we may obtain the magnetic field at about 28 Gs around the acceleration site. As the radio type II burst occurred above the upper HXR source region, the related magnetic field should be a bit of weaker than 25 Gs. Considering the uncertainty of Equation (2), it is reasonable to assume the magnetic field strength in the range of 15 - 40 Gs.

With the above results, furthermore, we may estimate the Alfven speed around the acceleration site and the source region of type II burst, $v_{A}=1650 - 4900$ km s$^{-1}$ (0.005 - 0.016 c). Considering the uncertainty of the estimated magnetic field strength, the Alfven speed is at least at the order of $v_{A}>1000$ km s$^{-1}$. It is much faster than the CME's velocity (about 600 km s$^{-1}$). This point is very important for understanding the physical processes underlying the above FDRBs.

\section{Physical Analysis}

Now that there are three different kinds of solar FDRBs: type III pair bursts, spike bursts, and spike-like bursts, then what are the physical processes underlying these bursts?

\subsection{Energy Estimation of the Nonthermal Electrons }

In order to understand the physical processes underlying the solar radio bursts, the first is to reveal the related emission mechanism and kinetic energy of the emitting electrons. The short lifetime, narrow frequency bandwidth and high brightness temperature indicate that all of the type III pairs, spikes and spike-like bursts should be produced by coherent emission processes. One of the possible candidates is electron cyclotron maser emission (ECME), a coherent process which is related to some magnetized plasma instabilities in relatively strong magnetic field, such as the loss-cone instability, etc. (Melrose \& Dulk 1982, Robinson 1991, Fleishman et al. 2003, Tang et al. 2012). However, ECME requires a relatively strong magnetic field which should exceed 35 - 146 Gs at the frequency of 100 - 410 MHz. As we mentioned in the above section, the magnetic field is only about 28 Gs around the acceleration site and below 25 Gs along the path of the type II burst. This fact indicates that it seems very difficult to excite ECME in this event.

The another candidate of the spikes and spike-like bursts is the coherent plasma emission (Zheleznyyakov \& Zlotnik 1975). With plasma emission and the observed $\bar{D}$, the speed of the emission source (Equation 3 in Tan 2013) can be estimated:

\begin{equation}
v_{s}\approx2H_{n}\bar{D}.
\end{equation}

Here, $H_{n}$ is the plasma barometric scale length. Generally, $H_{n}$ increases with the height of the emission source region above solar surface (Benz et al. 1983, Stahli \& Benz 1987, Aschwanden et al. 1995, etc.). However, as we do not know the exact height of the source region at certain frequency directly, it is still difficult to obtain the exact value of $H_{n}$. Here, we try to obtain an indirect estimation of $H_{n}$. Fig. 1 shows that the radio type III pairs occur around the peak time of HXR emission flux at energy of 50 - 100 keV, we may assume that the type III pairs-related energetic electrons have kinetic energy of about 50 - 100 keV, the corresponding velocities are in the range of 0.42 - 0.57 c. Substituting this velocity and the observed $\bar{D}$ into Equation (3), we may obtain an estimation $H_{n}\approx7.6\times10^{4}$ km. This value seems to be compatible to the other estimations at the frequency range of metric waves (Bean et al. 1983, Tan et al. 2016).

Using the above estimated $H_{n}$ and the observed $\bar{D}$ of radio spike bursts, then we may obtain an estimation of velocities of the emitting sources are about (1.7 - 4.0)$\times10^{4}$ km s$^{-1}$ (about 0.06 - 0.13 c) for the PD spikes and -(2.0 - 5.9)$\times10^{4}$ km s$^{-1}$ (about 0.07 - 0.2 c) for ND spikes. Obviously, these velocities are too fast to demonstrate the motions of CMEs or any solar plasma jets. They are possibly responsible for the moving of emitting nonthermal electron beams, and the corresponding kinetic energy is in the range of 0.8 - 4.4 keV for PD spikes and 1.1 - 10.0 keV for ND spikes, respectively.

Using the same method, we may estimate the velocities and kinetic energies of the emitting electrons associated to the spike-like bursts in the type II burst. As for ND spike-like bursts, the corresponding velocities and kinetic energies are (5.6 - 10.8)$\times10^{4}$ km s$^{-1}$ (about 0.19 - 0.36 c) and 9.1 - 35.6 keV, respectively. And for PD spike-like bursts, the corresponding velocities and kinetic energies are (6.2 - 10.2)$\times10^{4}$ km s$^{-1}$ (about 0.21 - 0.34 c) and 11.5 - 33.5 keV, respectively.

Although the above velocities of the nonthermal electrons are slower than that of the electron beams related to type III pair bursts, they are still much higher than that of the thermal elections in the background plasma and much higher than the Alfven speed in the background plasma. These nonthermal electrons are enough to trigger Langmuir waves and the related coherent plasma emission (Robinson \& Benz 2000).

\subsection{Acceleration Processes of the Nonthermal Electrons}

The electron acceleration is crucial for converting magnetic energy into kinetic energy in solar eruptions. However, so far, it is still remained uncertain which mechanism accelerates electrons and other charged particles. The existing competing mechanisms include acceleration by magnetic reconnection, turbulence, and shock waves (Miller et al. 1997, Tsuneta \& Naito 1998, Drake et al. 2006, Zharkova et al. 2011). Then, how do accelerate the nonthermal electrons related to the different kinds of FDRBs?

Fig. 7 shows our proposed physical explanation of the related electron accelerations and the generations of radio type III pair (normal and RS type III), spikes, and spike-like bursts in front of CME, respectively.

\begin{figure*} % Fig. 7
\begin{center}
   \includegraphics[width=10.0 cm]{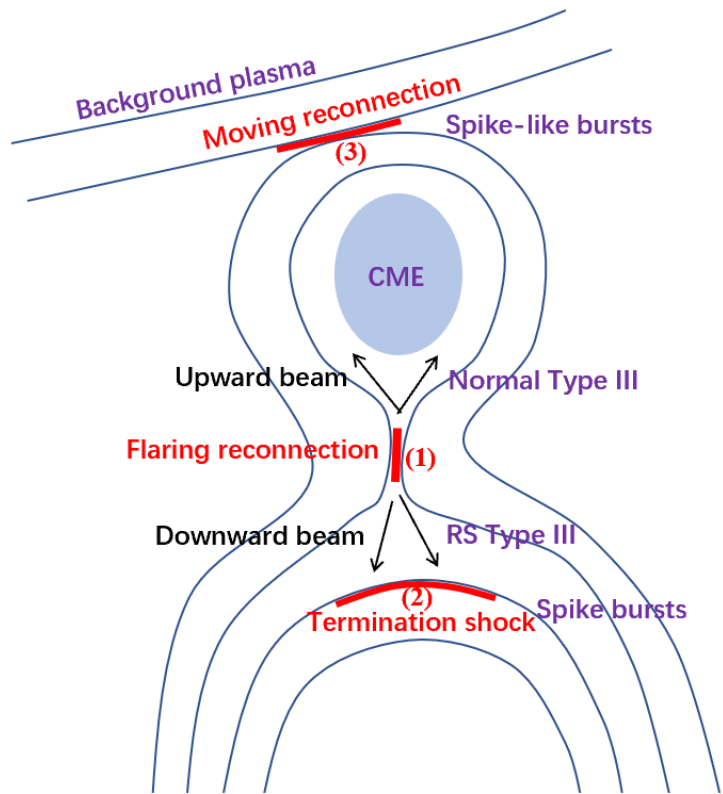}
\caption{A cartoon of showing the acceleration of flaring magnetic reconnection (1), termination shock wave (2) and the CME-driving moving reconnection (3), and the generation positions of radio type III pair bursts, spike bursts, and spike-like bursts, respectively. The thick red bands indicate the places of particle acceleration.}
\end{center}
\end{figure*}

(1) Type III pair bursts.

At first, it is naturally to suppose that the nonthermal electrons related to type III pair bursts should be accelerated by the flaring magnetic reconnection, which produce the upward nonthermal electron beams to generate the normal type III bursts and the downward nonthermal electron beams to generate the RS type III bursts. Therefore, the type III pair bursts are explained as produced by bi-directional electron beams from the flaring magnetic reconnection site (showing (1) in Fig. 7) and the separate frequency ($f_{x}$) may pinpoint the flare primary energy-release site where the magnetic reconnection and particle accelerations take place (Li et al. 2011, Tan et al. 2016). In this event, the separate frequency ($f_{x}$) is around 310 MHz and the corresponding plasma density is about 1.2$\times10^{15}$ m$^{-3}$. This separate frequency is relatively lower than that in most of the other flares (Aschwanden \& Benz 1997, Tan et al. 2016), and the corresponding height of the reconnecting site is around $H=$4.7$\times10^{4}$ km above the solar surface. As reported by Chen et al. (2016), the reconnection is classified as breakout type of magnetic reconnection.

(2) Spike bursts.

Considering the random distributions of the PD and ND spike bursts in big clusters, it is very possible that the related nonthermal electrons should be accelerated by some shock waves. Although there is a CME but it occurred after the spike bursts which implies that the CME has no relationship to the formation of the spike bursts. Additionally, the overall of spike bursts has no obviously frequency drifting rate, distinctly different from the type II burst with slowly drifting rates. The spike bursts just followed the type III pair bursts around the separate frequency, and this fact indicates that they should be related to the flaring magnetic reconnection. The above facts let us to suppose that the spike bursts should be produced by a group of nonthermal electrons accelerated by a shock wave which may be generated when the reconnecting fast outflows or downward high energy electron stream interact on the flaring loop top, similar to the flaring termination shock reported by Chen et al. (2015), showing (2) in Fig. 7.

(3) Spike-like bursts.

The formation of spike-like bursts should be related to the motion of CME. According to the traditional view, a fast CME may trigger a shock wave which accelerate the electrons to produce the type II radio burst and the spike-like bursts (Mann et al. 1995, Cane \& Erickson 2005). However, by adopting the estimation of $H_{n}$ and the observed frequency drifting rate of type II burst, we found that the speed of the emission source region ranges in 300 - 460 km s$^{-1}$. This speed is a bit slower than the measurement from EUV images (about 600 km s$^{-1}$, Chen et al. 2016) but still at the same order of magnitude. The speed is much slower than the Alfven speed around the acceleration site and the source region of type II burst, $v_{A}=1650 - 4900$ km s$^{-1}$ (0.005 - 0.014 c). The related CME is only a slow one which could not trigger a shock wave. Then what triggered these spike-like bursts? It is known that CME is a cloud of moving magnetized plasma, and the background coronal plasma is also permeated with magnetic field. Considering the distribution of the ND and PD spike-like bursts is very similar to the type III pair bursts which produced from magnetic reconnection, therefore we proposed that the interaction between CME and the background magnetized coronal plasma may generate moving magnetic reconnection (showing (3) in Fig. 7), and accelerate electrons to produce nonthermal energetic electrons which triggered the formation of type II radio burst and the spike-like bursts. The upward accelerated electrons from the moving magnetic reconnection may generate the ND spike-like bursts while the downward accelerated electrons from the moving magnetic reconnection will generate the PD spike-like bursts. The moving magnetic reconnection following a slow CME may explain the formation of the spike-like bursts superimposed on the type II radio burst naturally.

\section{Summary and Discussion}

Generally, people think that there is no obvious difference between solar radio spikes and spike-like bursts superimposed on type II burst for they should be generated from similar physical processes and mechanism. However, based on the careful parameter scrutinizing in this work, we find that they are actually distinctly different from each other. Actually in this work, we find that there are three kinds of FDRBs in an X-class flare. They are generated by nonthermal energetic electrons which accelerated by different physical processes.

(1) Type III pair bursts, broad bandwidth, fast frequency drifting rate, and the ND bursts occurred in the frequency below the separate frequency while the PD bursts occurred above the separate frequency. They are generated from the nonthermal electrons accelerated by the flaring magnetic reconnection.

(2) Spike bursts following the type III pair bursts, very short lifetime, very narrow bandwidth, and the ND and PD bursts are distributed randomly following the type III pairs around the separate frequency. They should be generated from the energetic electrons possibly accelerated by a termination shock wave above the flaring looptop. The shock wave is possibly formed when the reconnecting fast outflows or high energy electron stream impact on the flaring loop top.

(3) Spike-like bursts superimposed on the type II burst, also very short lifetime and narrow bandwidth, about 2 times bandwidth and nearly 2 times frequency drifting rates of spike bursts, the ND and PD bursts are distributed separately on two sides of the central ridges of the type II burst. They are generated from the energetic electrons possibly accelerated by a moving magnetic reconnection when the CME interacts on the background magnetized coronal plasma.

The above explanation sounds reasonable. However, it requires more observations to demonstrate its feasibility. Especially the termination shock wave acceleration to produce spike bursts and the moving magnetic reconnection to produce spike-like bursts need much more multiple observations to demonstrate the existence of special plasma loops, magnetic field configurations and radio bursts at the corresponding frequencies. These multiple observations should be including EUV images with high spatial resolution, broadband radio spectrometers with high temporal and frequency resolution, and spectral radio images at the corresponding frequency, such as the MUSER observations (Yan et al. 2009, Chen et al. 2019). In the near future, we plan to collect more flare events with multiple observations to investigate the relationships among the radio type III pair bursts, spike bursts, spike-like bursts, and the related flare and CME processes, statistically to reveal the real origin of nonthermal electrons from solar eruptions.

\acknowledgments The authors would thank Prof G. Chernov for helpful and valuable comments on this paper. The work is supported by NSFC Grants 11433006, 11573039, 11661161015, 11790301 and 11973057. We adopted radio observations from Iitate Planetary Radio Telescope (IPRT) which is a Japanese radio telescope developed and operated by Tohoku University, UV and EUV observation data was obtained from AIA/SDO. This work was also partly supported by the international joint research program of the Institute for Space-Earth Environmental Research at Nagoya University and JSPS KAKENHI, Grant No. 18H01253 and the International Space Science Institute-Beijing (ISSI-BJ).


\begin{thebibliography}{}
\bibitem[Achwanden(1993)]{Achwanden1993}Achwanden, M. J., Benz, A.O., Schwartz, R.A.: 1993, \emph{ApJ}, \textbf{417}, 790
\bibitem[Achwanden(1995)]{Achwanden1995}Achwanden, M. J., Benz, A.O., Dennis, B.R., \& Schwartz, R.A.: 1995, \emph{ApJ}, \textbf{455}, 347
\bibitem[Achwanden(1997)]{Achwanden1997}Achwanden, M. J., Benz, A.O.: 1997, \emph{ApJ}, \textbf{480}, 825
\bibitem[Armatas(2019)]{Armatas2019}Armatas, S., Bouratzis, C., Hillaris, A., Alissandrakis, C.E., Preka-Papadema, P., Moussas, X., et al.: 2019, \emph{A$\&$A}, \textbf{624}, A76
\bibitem[Bastian(1998)]{Bastian1998}Bastian, T.S., Benz, A.O., \& Gary, D.E.: 1998, \emph{Annu. Rev. Astron. Astrophys.}, \textbf{36}, 131
\bibitem[Benz(1982)]{Benz1982}Benz, A.O., Zlobec, P., \& Jaeggi, M.: 1982, \emph{A$\&$A}, \textbf{109}, 305
\bibitem[Benz(1983)]{Benz1983}Benz, A.O., Bernold, T.E.X., \& Dennis, B.R.: 1983, \emph{ApJ}, \textbf{271}, 355
\bibitem[Benz(1985)]{Benz1985}Benz, A.O.: 1985, \emph{SoPh}, \textbf{96}, 357

\bibitem[Cane(2005)]{Cane2005}Cane, H. V., Erickson, W. C.: 2005, \emph{ApJ}, \textbf{623}, 1180
\bibitem[Chernov(2010)]{Chernov2010}Chernov, G.P., Yan Y.H., Tan C.M., Chen B., Fu Q.J.: 2010, \emph{SoPh}, \textbf{262}, 149
\bibitem[Chen(2015)]{Chen2015} Chen, B., Bastian, T.S., Shen, C.C., Gary, D.E., Krucker, S., Glesener, L.: 2015, \emph{Science}, \textbf{350}, 1238
\bibitem[Chen(2016)]{Chen2016} Chen, Y., Du, G.H., Zhao, D., Wu, Z., Liu, W., Wang, B., et al.: 2016, \emph{ApJL}, \textbf{820}, L37
\bibitem[Chen(2019)]{Chen2019} Chen, X. Y., Yan, Y.H., Tan, B. L., Huang, J., Wang, W., Chen, L. J., et al.: 2019, \emph{ApJ}, \textbf{878}, 78
\bibitem[Drake(2006)]{Drake2006}Drake, J.F., Swisdak, M., Che, H., Shay, M.A.: 2006, \emph{Nature}, \textbf{443}, 553
\bibitem[Dulk(1971)]{Dulk1971}Dulk, G.A., Altschuler, M.D.: 1971, \emph{SoPh}, \textbf{20}, 438
\bibitem[Dulk(1978)]{Dulk1978}Dulk, G.A., McLean, D.J.: 1978, \emph{SoPh}, \textbf{57}, 279
\bibitem[Dulk(1985)]{Dulk1985}Dulk, G.A.: 1985, \emph{Annu. Rev. Astron. Astrophys.}, \textbf{23}, 169
\bibitem[Fleishman(2003)]{Fleishman2003} Fleishman, G.D., Gary, D.E., \& Nita, G.M.: 2003, \emph{ApJ}, \textbf{593}, 571
\bibitem[Iwai(2012)]{Iwai2012}Iwai, K., Tsuchiya F., Morioka, A., Misawa, H.: 2012, \emph{SoPh}, \textbf{277}, 477

\bibitem[Lemen(2012)]{Lemen2012}Lemen, J. R., Title, A. M., Akin, D. J., et al.: 2012, \emph{SoPh}, \textbf{275}, 17
\bibitem[Lin(1971)]{Lin1971}Lin, R. P., \& Hudson, H. S.: 1971, \emph{SoPh}, \textbf{17}, 412
\bibitem[Lin(1981)]{Lin1981}Lin, R. P., Potter, D. W., Gurnett, D. A., \& Scarf, F. L.: 1981, \emph{ApJ}, \textbf{251}, 364
\bibitem[Lin(2002)]{Lin2002}Lin, R. P., Dennis, B. R., Hurford, G. J., Smith, D. M., Zehnder, A., Harvey, P.R., et al.: 2002, \emph{SoPh}, \textbf{210}, 3
\bibitem[Mann(1995)]{Mann1995}Mann, G., Cla$\beta$en, T., Aura$\beta$, H.: 1995, \emph{A$\&$A}, \textbf{295}, 775
\bibitem[Melrose(1982)]{Melrose1982} Melrose, D. B., Dulk, G.A.: 1982, \emph{ApJ}, \textbf{259}, 844
\bibitem[Miller(1997)]{Miller1997}Miller, J. A., Cargill, P. J., Emslie, A. G., Holman, G. D., Dennis, B.R., LaRosa, T.N., et al.: 1997, \emph{J. Geophys. Res.}, \textbf{102}, 14631
\bibitem[Ning(2000)]{Ning2000}Ning, Z. J., Fu, Q.J., \& Lu, Q. K.: 2000, \emph{SoPh}, \textbf{194}, 137
\bibitem[Reid(2014)]{Reid2014} Reid, H. A. S., \& Ratcliffe, H.: 2014, \emph{RAA}, \textbf{14}, 773
\bibitem[Robinson(2000)]{Robinson2000}Robinson, P.A., Benz, A.O.: 2000, \emph{SoPh}, \textbf{194}, 345
\bibitem[Robinson(1991)]{Robinson1991}Robinson, P.A.: 1991, \emph{SoPh}, \textbf{134}, 299
\bibitem[Stahli(1987)]{Stahli1987}Stahli, M., \& Benz, A.O.: 1987, \emph{A$\&$A}, \textbf{175}, 271
\bibitem[Tan(2010)]{Tan2010} Tan, B. L., Zhang, Y., Tan, C.M., Liu, Y.Y.: 2010, \emph{ApJ}, \textbf{723}, 25
\bibitem[Tan(2013)]{Tan2013} Tan, B. L.: 2013, \emph{ApJ}, \textbf{773}, 165
\bibitem[Tan(2014)]{Tan2014} Tan, B. L., Tan, C.M., Zhang, Y., Meszarosova, H., Karlicky, M.: 2014, \emph{ApJ}, \textbf{780}, 129
\bibitem[Tan(2016)]{Tan2016} Tan, B.L., Karlicky, M., Meszarosova, H. \& Huang, G.L.: 2016, \emph{RAA}, \textbf{16}, 74
\bibitem[Tang(2012)]{Tang2012} Tang, J.F., Wu, D.J., Yan, Y.H.: 2012, \emph{ApJ}, \textbf{745}, 134
\bibitem[Tsuneta(1998)]{Tsuneta1998} Tsuneta, S., Naito, T.: 1998, \emph{ApJ}, \textbf{495}, L67
\bibitem[Yan et al.(2009)]{Yan2009} Yan, Y. H., Zhang, J., \& Wang, W., et al.: 2009, \emph{EM$\&$P}, \textbf{104}, 97
\bibitem[Zharkova(2011)]{Zharkova2011} Zharkova, V. \& Siversky, T.: 2011, \emph{ApJ}, \textbf{733}, 33
\bibitem[Zheleznyakov(1975)]{Zheleznyakov1975}Zheleznyakov, V.V., Zlotnik, E.Y.: 1975, \emph{SoPh}, \textbf{44}, 461
\end{thebibliography}
\end{document}